\documentclass[11pt]{article}
\usepackage[utf8]{inputenc}
\usepackage{amsmath}
\usepackage{hyperref}
\usepackage{graphicx}
\usepackage{caption}
\usepackage{subcaption}

\title{Nonparametric clustering of RNA-sequencing data}
\author{Gabriel Lozano \thanks{Universidad Nacional de Colombia, golozanop@unal.edu.co} \and Nadia Atallah\thanks{Department of Comparative Pathobiology, Purdue Univesity, natallah@purdue.edu} \and Michael Levine\thanks{Department of Statistics, Purdue University, mlevins@purdue.edu}}

\begin{document}
\newcommand{\diff}{\mathrm{d}}
\newcommand{\op}[1]{\mathcal{#1}}
\newcommand{\p}{\pi} 
\newcommand{\marginals}{\varphi}
\newcommand{\copula}{\theta}
\newcommand{\vv}[1]{\mathbf{#1}} 
\newcommand{\alert}[1]{\textcolor{red}{#1}}
\newcommand{\eps}{\varepsilon}
\newcommand{\truth}[1]{#1^\ast}
\newcommand{\init}[1]{#1}
\newcommand{\minimizer}[1]{\widehat{#1}}
\newcommand{\norm}[1]{\left\lVert #1 \right\rVert}

\maketitle

\abstract{Identification of clusters of co-expressed genes in transcriptomic data is a difficult task. Most algorithms used for this purpose can be classified into two broad categories: distance-based or model-based approaches. Distance-based approaches typically utilize a distance function between pairs of data objects and group similar objects together into clusters. Model-based approaches are based on using the mixture-modeling framework. Compared to distance-based approaches, model-based approaches offer better interpretability because each cluster can be explicitly characterized in terms of the proposed model. However, these models present a particular difficulty in identifying a correct multivariate distribution that a mixture can be based upon. In this manuscript, we review some of the approaches used to select a distribution for the needed mixture model first. Then, we propose avoiding this problem altogether by using a nonparametric MSL (Maximum Smoothed Likelihood) algorithm. This algorithm was proposed earlier in statistical literature but has not been, to the best of our knowledge, applied to transcriptomics data. The salient feature of this approach is that it avoids explicit specification of distributions of individual biological samples altogether, thus making the task of a practitioner easier. When used on a real dataset, the algorithm produces a large number of biologically meaningful clusters and compares favorably to the two other mixture-based algorithms commonly used for RNA-seq data clustering. Our code is publicly available in Github at 
\url{https://github.com/Matematikoi/non_parametric_clustering}
}


\section{Introduction}

Increasingly complex studies of transcriptome dynamics can be carried out now using high-throughput sequencing of reverse-transcribed RNA molecules. Such a procedure is typically called RNA-sequencing (RNA-seq). Studying RNA-seq data helps researchers gain a deeper understanding of how changes in transcriptional activity reflect various cell types and contribute to phenotypic differences. One of the ways to gain new insights from RNA-seq data is to identify groups (clusters) of co-expressed genes. This can help researchers target genes involved in similar biological processes, thus helping in the design of new pharmaceutical drugs, or finding genes that are candidates for co-regulation. Identification of clusters of co-expressed genes also helps characterize biological functions for orphan genes. 

By now, a number of algorithms has been proposed for clustering RNA-seq data. Note that these data have a number of characteristics that makes modeling them rather difficult. First of all, they tend to be highly skewed and have a large dynamic range. Second, they typically demonstrate positive correlation between the gene length and read counts. Third, these data are almost always overdispersed (i.e. their variance is larger than their mean).

Most of the model-based clustering methods used for RNA-seq data typically view each cluster as represented by a distinct distribution, while the entire dataset is modeled as a finite mixture of these distributions. One of the advantages of this approach is that it allows a researcher to assess the appropriate number of clusters, the distance between clusters, and test hypotheses about these quantities. In the existing literature, most of these models tend to be parametric; in other words, it is assumed that distributions of individual biological samples are modeled as belonging to a particular distribution. The choice of distribution ranges from Poisson \cite{rau2015co} to negative binomial \cite{si2014model} to complicated distributions built specifically for the given task, such as multivariate Poisson-lognormal \cite{silva2019multivariate,subedi2020parsimonious}. The choice that has to be made here is a rather difficult one as there are few ready-made goodness of fit tests for most multivariate discrete distributions. As a practical concern, many practitioners may find it difficult to make such a choice in practice. Thus, an approach that avoids making such a choice altogether will be beneficial. 

In this manuscript, we suggest an alternative way of handling the uncertainty in choosing a distribution for biological samples. Instead of trying to make a choice between various parametric families, we suggest viewing these distributions as ``just" probability density functions that do not belong to any such family. The nonparametric approach to multivariate mixture modeling and clustering has been studied in statistics for some time; several methods have been proposed to fit such models and establish the structure of relevant clusters \cite{benaglia2009like, levine2011maximum, chauveau2015semi}. However, their use in bioinformatics in general has been very limited (see e.g. \cite{anchang2014ccast}); to the best of our knowledge, they have not been used to cluster the data at all. 

In this work, we suggest the use of an algorithm that can fit a general multivariate nonparametric mixture model with conditionally independent marginals in the RNA-seq data context. This method is a so-called npMSL (the nonparametric Maximum Smoothed Likelihood) method that was originally proposed in \cite{levine2011maximum}. The corresponding algorithm is an MM (Maximization-Minorization) algorithm that possesses the monotonicity property, similarly to the EM algorithm, and is guaranteed to converge. 

The remainder of this article is organized as follows. In the Methods section, we describe the model used in detail, introduce the necessary notation, and define the algorithm that is to be used to perform the clustering task. Here, we also describe the model selection procedure that we use to select an appropriate number of clusters. In the Real Data Analysis section, we describe in detail two datasets that we use to illustrate our approach. The results are illustrated using both a simple visualization method and the results of a GO functional enrichment analysis. Finally, some concluding remarks are provided in the Discussion section.

\section{Methods}

Let $Y_{ij}$  be the random variable corresponding to the digital gene expression measure for a biological entity $i$ ($i=1,\ldots,n)$ of condition $j$ $(j=1,\ldots, d).$ We denote the corresponding observed value $y_{ij}.$ This setting implies that the data $\vv y$ is the $n\times d$ matrix of the digital gene expression for all observations and variables. Also, $\vv y_{i}$ is the $d-$ dimensional vector of digital gene expression for all variables of observation $i.$

\subsection{Nonparametric mixture model with conditionally independent measurements}
Historically, it has been common to use parametric mixture models to cluster RNA-seq data (see e.g.\cite{si2014model,rau2015co,rau2018transformation}). To the best of our knowledge, nonparametric approach to model-based clustering of RNA-seq data has not been tried before. At the same time, it has been used in many other statistical application areas e.g. developmental psychology, hydrology, and others \cite{chauveau2015semi, shen2018mm}. This approach assumes that the density functions of clusters do not come from a particular parametric family (e.g. Gaussian, Poisson etc.). More specifically, 
the data are assumed to come from $m$ distinct clusters with the density of the $k$th cluster, $k=1,\ldots,m$ being $f_{k}.$ It is further assumed that each one of these densities $f_{k}$ is equal with probability $1$ to the product of its marginal densities:
\begin{equation}\label{marg_independence}
f_{k}({\vv y_i})=\prod _{j=1}^{d}f_{kj}(y_{ij})
\end{equation}
with $\vv y_i=(y_{i1},\ldots,y_{id})^{'}.$ Taking a fully nonparametric approach with regard to the $f_{kj}$, we may therefore express the density function of any observation $\vv Y_i$ according to a nonparametric finite mixture model as
\begin{equation}\label{model}
\vv Y_i \ \sim \ 
g_{\boldsymbol\theta}({\vv y}_{i})\ =\ \sum_{k=1}^{m}\pi_{k}\prod_{j=1}^{d}f_{kj}(y_{ij}),
\end{equation}
where $\boldsymbol\pi=(\pi_{1},\ldots,\pi_{m})$ must satisfy
\begin{equation}\label{lambdaconstraints}
\sum_{k=1}^m \pi_k=1 \quad\mbox{and each $\pi_k\ge 0$.}
\end{equation}
Here, we assume $\vv Y_i = (Y_{i1}, \ldots, Y_{id})^{'}$ and we let
 $\boldsymbol\theta$ denote the vector of parameters to be estimated,
 including the mixing proportions $\pi_{1},\ldots,\pi_{m}$ and the
univariate densities $f_{kj}$.  For convenience, we will also use the notation $\vv f$ for a vector of all the marginal densities $\{f_{kj}\},$ $j=1,\ldots,d,$ $k=1,\ldots,m.$ Furthermore, throughout this article, $k$ and $j$ always denote the component and coordinate indices, respectively;
thus, $1\le j\le d$ and $1\le k \le m.$ Thus, the overall population of the sample size $n$ is distributed according to 
\begin{equation}\label{prod_marg}
f(\vv y_i; m,\boldsymbol\theta)=\prod_{i=1}^{n}\sum_{k=1}^{m}\pi_{k}\prod_{j=1}^{d}f_{kj}(y_{ij}).
\end{equation}

The conditional independence may seem rather limiting at first sight. However, it may be thought of as a simplification of the commonly used repeated measures random effects model. In such a model, one usually assumes that the multivariate observations on an individual are independent, conditional on the identity of the individual in question. Here, the individual-level effects are replaced by the entity (gene)- level effects. Note also that the conditional independence assumption has been used on a number of occasions when modeling RNA-seq data using parametric mixtures e.g. \cite{rau2015co,si2014model}. On the other hand, no individual marginal density $f_{jk}$ is assumed to have come from a family of densities indexed by a finite-dimensional parameter vector, such as Gaussian, Student, etc. Such a nonparametric approach represents a substantial generalization compared to approaches typically used in bioinformatics literature. 

\subsection{Inference}

We are going to start with some needed notation. First, let $Z_{ik}\in \{0,1\}$ be a Bernoulli random variable indicating that an individual $i$ comes from the component $k.$  Because each individual comes from exactly one component, it implies that $\sum_{k=1}^{m}Z_{ik}=1.$ Therefore, the complete data would be the set of all $(\vv x_i,\vv Z_i),$ where $1\le i \le n$. This suggests that the entire data can be viewed as consisting of observable and unobservable parts. Therefore, an EM-type algorithm seems to be an appropriate option for estimating of parameters of the model \eqref{model}. 

\cite{levine2011maximum} introduced an algorithm that minimizes a smoothed loglikelihood function of the data produced by the model \eqref{model}. This algorithm has a provable monotonicity property. We only give a brief description of this algorithm; for detailed discussion of its monotonicity property see \cite{levine2011maximum}. For the purpose of this discussion, some additional notation is needed. 

Let $K(\cdot)$  be a kernel density function on the real line. With a slight abuse of notation, let us
define the product kernel function in the $d$-dimensional space as $K({\vv u})=\prod_{j=1}^{d}K(u_{j})$ and its rescaled version $K_{h}({\vv u})=h^{-d}\prod_{j=1}^{d}K(h^{-1}u_{j})$ for a positive parameter $h$ that is commonly called the bandwidth. Furthermore, we smooth a function $f$ using the following smoothing operation ${\mathcal S}f(\vv y)=\int K_h(\vv y-\vv u)f(\vv u)\,d\vv u.$ The same smoothing operation can be applied to an $m$-dimensional vector of functions by defining ${\mathcal S}\vv f = ({\mathcal S}f_1, \ldots, {\mathcal S}f_m)^{'}.$  We also define a nonlinear smoothing operation ${\mathcal N}$ as
\[
{\mathcal N} f(\vv y)\ =\ \exp\left\{ ( {\mathcal S} \log f )(\vv y) \right\}
\ =\ \exp \int K_h(\vv y- \vv u) \log f(\vv u) \, d\vv u.
\]

To simplify notation, we introduce the finite mixture operator ${\mathcal M}_{\boldsymbol\pi}\vv f(\vv y):=\sum_{k=1}^{m}\pi_{k}f_{k}({\vv y}),$ whence we obtain ${\mathcal M}_{\boldsymbol\pi}\vv f(\vv y) =g_{\boldsymbol\theta}(\vv y).$ Also, we denote ${\mathcal M}_{\boldsymbol\pi} {\mathcal N}\vv f(\vv y):=\sum_{k=1}^{m}\pi_{k}{\mathcal N}f_{k}({\vv y}).$ With this notation in mind, we define the following algorithm. Given initial values $(\vv f^0, \boldsymbol\pi^0),$ iterate the following three steps for $t=0,1,\ldots:$
\begin{itemize}
\item{\bf E-step:\ }
Define, for each $i$ and $k$,
\begin{equation}\label{wij}
w_{ik}^t
= \frac{\pi^t_k {\mathcal N} f_k^t (\vv y_i) } {{\mathcal M}_{\boldsymbol\pi^t}
{\mathcal N}\vv f^t (\vv y_i) }
= \frac{\pi^t_k {\mathcal N} f_k^t (\vv y_i) } {\sum_{a=1}^m
\pi_a^{t} {\mathcal N}f_{a}^t (\vv y_i) }.
\end{equation}
\item
{\bf M-step, part 1:\ }
Set
\begin{equation}\label{lambda}
\pi_k^{t+1} = \frac{1}{n}\sum_{i=1}^n w_{ik}^t
\end{equation}
for $k=1, \ldots, m$.
\item
{\bf M-step, part 2:\ }
For each $j$ and $k$, let
\begin{align}\label{densest}
&f_{kj}^{t+1}(u) = \frac { \sum_{i=1}^n w_{ik}^t
K_h\left(u-y_{ij}\right) }
{\sum_{i=1}^n w_{ik}^t  }
\\
&=\frac {1}{nh\pi_{k}^{t+1}}  \sum_{i=1}^n w_{ik}^t
 K\left(\frac{u-y_{ij}}{h}\right)\nonumber.
\end{align}
\end{itemize}
Let us define the following functional of $\boldsymbol\theta$ (and, implicitly, $g$): 
\begin{equation}\label{elldefn}
\ell(\boldsymbol\theta)=\int g({\vv y})\log\frac{g({\vv y})}{[{\mathcal M}_{\boldsymbol\pi}{\mathcal N}\vv f]({\vv y})}
\,d{\vv y}.
\end{equation}
This functional represents conceptually a smoothed negative log-likelihood. Then, \cite{levine2011maximum} shows that the value of this functional decreases at each step of the introduced algorithm. This algorithm will be referred to as npMSL (nonparametric Maximum Smoothed Likelihood) algorithm.

The npMSL algorithm can also be generalized to a model where there are blocks of coordinates that are identically distributed (in addition to being conditionally independent). If we let $b_{j}$ be the block index of the $j$th coordinate, where $1\le b_{j}\le L$ and $L$ is the total number of such blocks, then the model \eqref{model} is modified as 
\[
g_{\boldsymbol\theta}({\vv y}_{i})\ =\ \sum_{k=1}^{m}\pi_{k}\prod_{j=1}^{d}f_{kb_{j}}(y_{ij}).
\]
If all the blocks have the size $1,$ we are back to the original model \eqref{model}. The nonlinear smoothing operator ${\mathcal N}f_{k}=\prod_{j=1}^{d}{\mathcal N}f_{kb_{j}}$ and definitions of ${\mathcal M}_{\boldsymbol\pi}\vv f$ and ${\mathcal M}_{\boldsymbol\pi}{\mathcal N}\vv f$ remain unchanged. The only element of the algorithm that actually needs an update is the density estimation step \eqref{densest}. For the $k$th component and block $l\in \{1,\ldots,L\},$ we now have
\[
f_{kl}^{t+1}(u)=\frac{\sum_{j=1}^{d}\sum_{i=1}^{n}w_{ik}^{t}I_{b_{j}=l}K_{h}(u-y_{ij})}{\sum_{j=1}^{d}\sum_{i=1}^{n}w_{ik}^{t}I_{b_{j}=l}}
\] 
where $I_{b_{j}=l}$ is an indicator function of the event $b_{j}=l$. 
The block version of the npMSL algorithm is the one that we have used in our study. In other words, when applying the npMSL method, we assume that replicates of the same condition represent a block of identically distributed coordinates.


\subsection{Bandwidth and kernel function selection issues}

Of the two issues - kernel and bandwidth selection - mentioned in the heading above, the first is a simpler of the two. There seems to be a general consensus in the literature on the density estimation that the choice of the kernel function does not matter much, at least in terms of efficiency of resulting estimators; see e.g. \cite{scott2015multivariate} for a general discussion of this issue. 

On the contrary, sensible choice of the bandwidth $h$ is a challenging problem.  The form of the npMSL algorithm introduced here assumes that $h$ is the same for each component and coordinate, or block. It is straightforward to introduce component- and block-specific bandwidths $h_{jl}$. Note also that individual component densities are not observed in the mixture setting. This fact complicates selection of the bandwidth in the mixture setting.


From the practical viewpoint, we found out that selecting a constant bandwidth according to the so-called Silverman's rule of thumb \cite{silverman2018density}, p. 48, works reasonably well. This method suggests choosing the bandwidth as 
\begin{equation}\label{Silverman}
h=0.9(nd)^{-1/5}\min\left\{SD,\frac{IQR}{1.34}\right\}
\end{equation}
where $SD$ is the standard deviation, $IQR$ the interquartile range, and $nd$ is the size of the entire dataset. 

Note that this is a rather crude method in the nonparametric mixture
setting. It is possible that it may result in under- or oversmoothing. First, pooling all of the data implies that one treats all of the mixture components as though they are from the same distribution. This can lead to an inflation of the bandwidth, especially if the mixture components’
centers are well-separated. This is true because, in such a case, the variability of the pooled
dataset will be larger than that of the individual components. Similarly, if
the vector coordinates are not identically distributed within each component/block,
the bandwidth could be biased upward for the same reason.
Yet operating in the opposite direction is the fact that the expression $nd$ in the formula defining the bandwidth above is an overestimate of the ``true" sample size. One can think of the ``true" sample size from each component being approximately equal to $\lambda_{k}nd$. 

The arguments above show first of all that it would be useful to know
something about the mixture structure in order to select a bandwidth. This
suggests an iterative procedure in which the value of $h$ is modified, and
the algorithm reapplied, after the output from the algorithm is obtained. This, however, is going to result in the violation of monotonicity property of npMSL algorithm. For a more detailed discussion of this topic, see \cite{benaglia2011bandwidth}. As the above suggests, a careful exploration of the bandwidth selection question is a research topic unto itself. Thus, to make our application of npMSL algorithm to the analysis of RNA-seq data simpler, we are only using the constant bandwidth value selected according to \eqref{Silverman}. 

\subsection{Selecting the number of clusters}\label{Model_Selec}

The npMSL algorithm assumes that the number of clusters is known in advance. This is almost never true when working with biological data. Unlike the case of parametric mixture models, there are few if any approaches to determining the number of clusters in a nonparametric mixture model. Some preliminary results in this direction have been obtained in \cite{kwon2021estimation}. We suggest the following approach to this problem. Recall that our nonparametric approach to clustering implies, as a first step, fitting a repeated measures model \eqref{model}. In light of this, it seems reasonable to fit first a {\it Gaussian} repeated measures mixture model over a range of possible number of clusters. To do this, we use the function repnormmix.sel of the mixtools package \cite{benaglia2009mixtools}. For every choice of the number of clusters, the initial cluster means were generated from a multivariate normal distribution of the correct dimensionality with the diagonal covariance matrix. We used the default settings of the function repnormmix.sel whereby the mean is determined by a normal distribution according to a binning method done on the data while the reciprocal of the variance has random standard exponential entries also according to a binning method done on the data. The initial probability weights were generated from a uniform Dirichlet distribution. For more details see the package description at \cite{benaglia2009mixtools}. The procedure used generated values of four model selection criteria - Akaike’s information criterion (AIC), Schwartz’s Bayesian information criterion (BIC), Bozdogan’s consistent AIC (CAIC), and Integrated Completed Likelihood (ICL) in order to choose the optimal number of clusters \cite{bozdogan1987model,biernacki2000assessing}.  In cases where different criteria suggested different choices for the number of clusters in a model, we used the choice suggested by at least two (out of the four) criteria. We have not experienced a situation where all four of the criteria suggested different choices for the number of clusters.

\section{Real data analysis}

In the following, we illustrate the use of npMSL algorithm using two real RNA-seq datasets. Note that it is not possible to compare the co-expression results obtained using these two methods to a ``true" clustering of the data, as, in general, such a classification does not exist. In order to identify whether the co-expressed genes seem to be implicated in similar biological processes, we conduct functional enrichment analysis of gene ontology (GO) terms for the clusters identified by the suggested methods. The data that we are using are a mouse RNA-seq dataset consisting of lung, kidney, liver, and small intestine tissues  \cite{Tsoucas:2019tx} and a prostate cancer cell-line RNA-seq dataset \cite{liu2019PCa}. The data are written in the matrix form where each row corresponds to a gene and each column to an experimental condition. The row names are the ENSEMBL gene names for each gene (ENSEMBL is a genome database project that is a scientific project of the European Bioinformatics Institute). Each row constitutes a digital gene expression of a particular gene across a set of cell lines (together with replicates) in our case. The goal is to cluster digital gene expression profiles in order to discover networks of co-expressed genes. 

Unless explicitly stated to the contrary, both datasets are normalized first. The normalization procedure that we use is commonly called FPKM (Fragments Per Kilobase of transcript per Million mapped reads). Let us denote the result of this procedure $Y_i$ for the $i$th gene. Then, it is defined as 
\[
Y_i = \frac{X_i}{N*s_i}*10^9
\]
where $N$ is the total number of reads sequenced, $s_i$ is the length of gene i, and $X_i$ is the number of counts for the $i$th gene. This type of normalization is commonly used for visualization and clustering. It is necessary if we want to be able to perform within sample comparisons (“gene A is expressed higher or lower than gene B”) because it is, effectively, a procedure that normalizes for gene length. Such a normalization is in order because, the longer the gene’s length is, the more fragments (“reads”) we sequence from that gene. 

The data sets have also been filtered using a cutoff of 1.5 CPM (counts per million), in which gene read counts are divided by the sum of read counts for a given sample  and multiplied by a million.  The CPM of the $i$th gene, denoted as $CPM_i$ is defined as
\[
CPM_i = \frac{X_i}{N}*10^6
\]
 where $N$ is the total number of reads sequenced for a given sample and $X_i$ is the number of counts for the $i$th gene.  

\subsection{Mouse tissue dataset}
The first dataset that we focus on is a dataset in which bulk RNA-seq was performed to profile the gene expression profiles in lung, kidney, liver, and small intestine tissues from six- to ten-week-old male C57BL/6J mice.  As gene expression data is highly tissue-specific, data such as these can allow for the identification of tissue-specific expression modules.  Two biological replicates are present for each tissue type and separate mice were used for each replicate.   Observations in these datasets are commonly referred to as ``counts" where a count is the number of reads that align to a particular feature(gene).  Filtering at the level of $1.5$ CPM as described above resulted in a file containing $116,512$ genes.  These data are available through the Gene Expression Omnibus (GEO) repository through accession number GSE124419. 

It has been also  noted earlier that the RNA-seq data often have a very large dynamic range and tend to be heavily skewed. This often represents a significant problem when modeling these data. Therefore, before modeling the data, we applied a logarithmic transform to it as a first step.

To select an appropriate number of clusters, we fit several Gaussian repeated measures models over a range of cluster numbers from $1$ to $20,$ as described earlier (in more details) in the Section \eqref{Model_Selec}. The comparison of the four model selection criteria suggests using the majority vote criterion that $17$ is an optimal number of clusters for the npMSL method. In this case, the procedure was not sensitive to the choice of starting values.

\begin{figure}
\centering
     \begin{subfigure}[t]{0.3\textwidth}
         \centering
         \includegraphics[width=\textwidth]{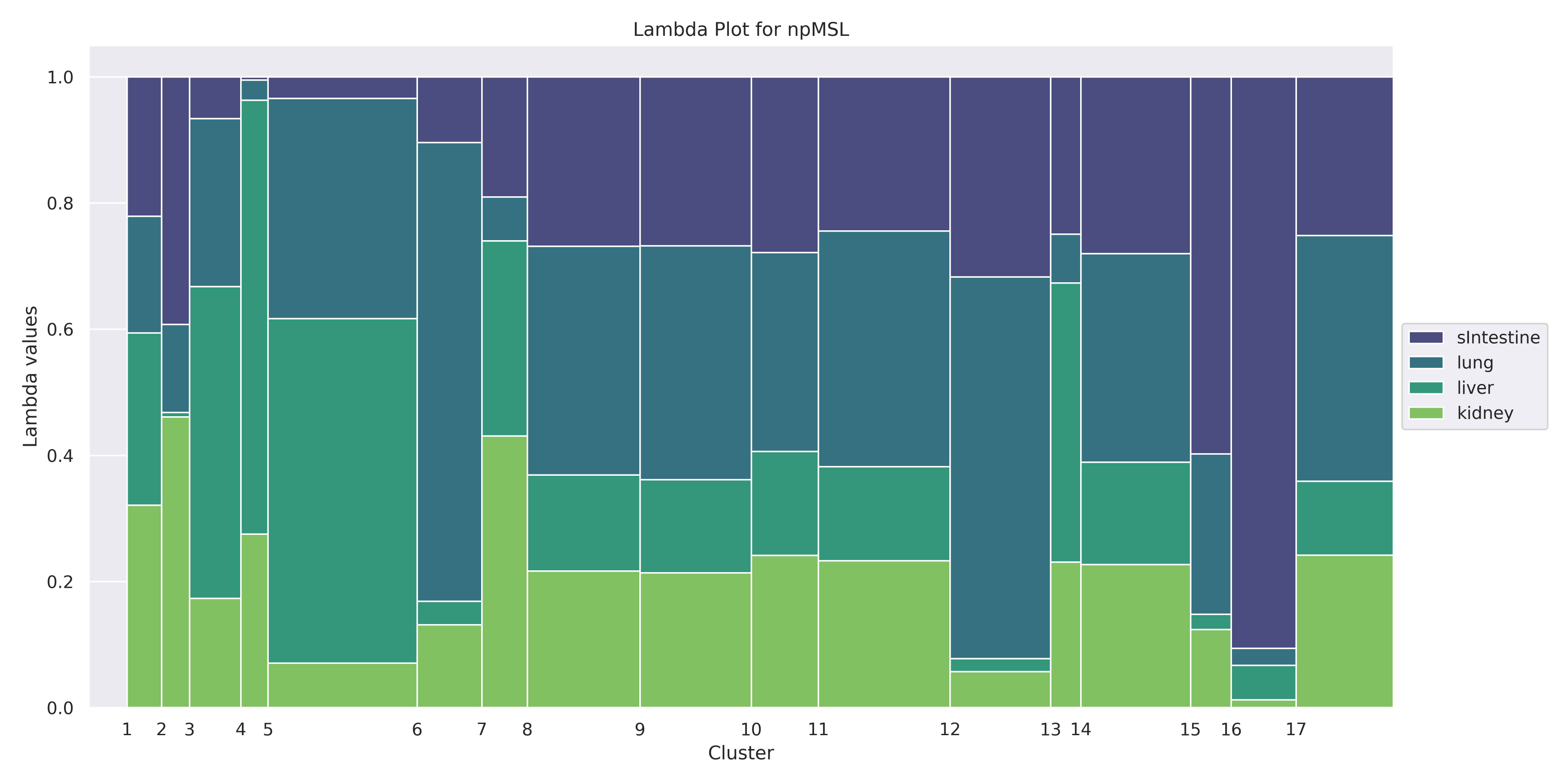}
         \caption{Cluster behavior for the mouse tissue dataset:npMSL}
         \label{fig:npMSL_mouse}
     \end{subfigure}
    \hfill
\begin{subfigure}[t]{0.3\textwidth}
         \centering
         \includegraphics[width=\textwidth]{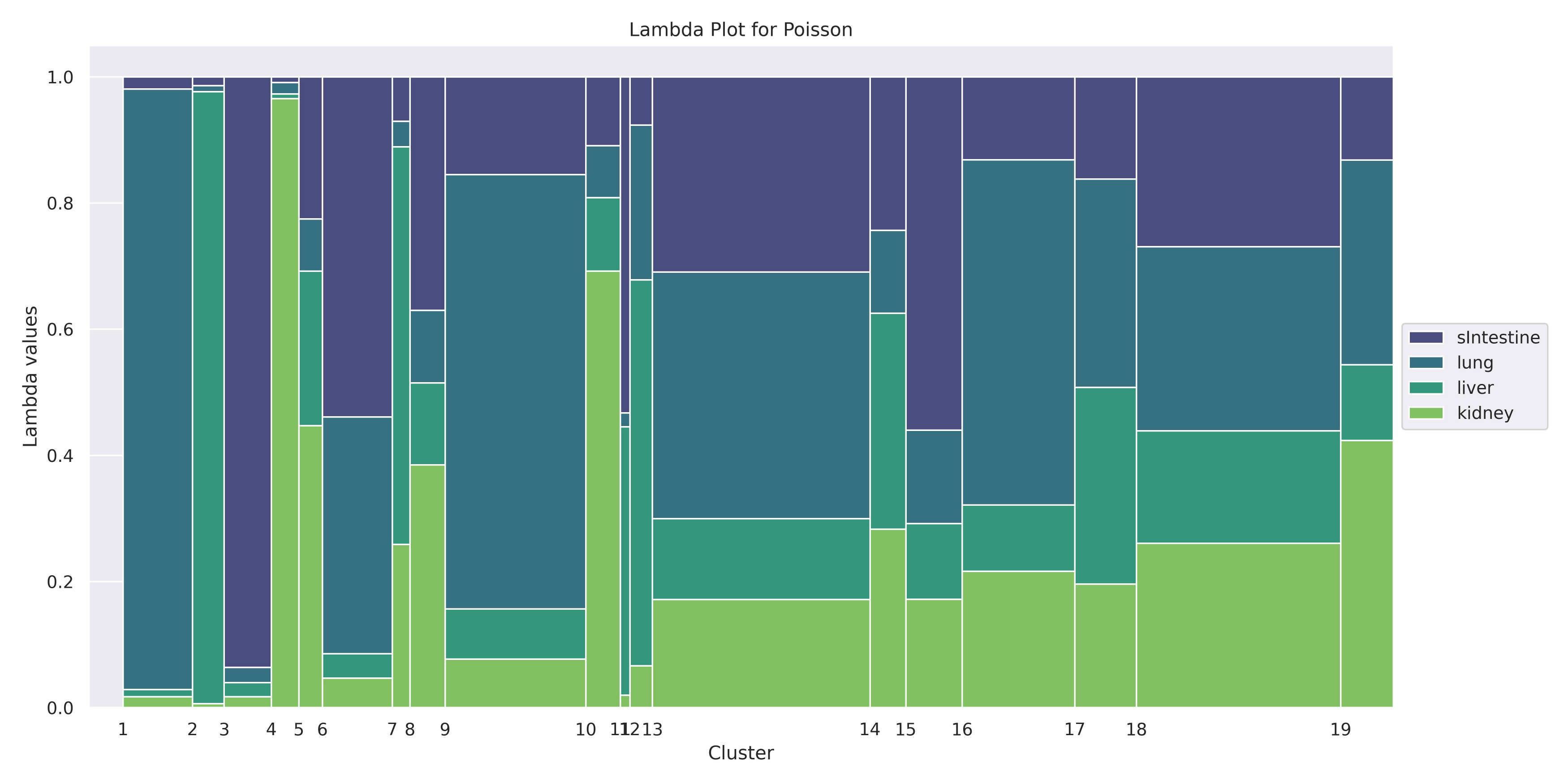}
         \caption{Cluster behavior for the mouse tissue dataset:Poisson method}
         \label{fig:Poisson_mouse}
     \end{subfigure}
     \hfill
\begin{subfigure}[t]{0.3\textwidth}
         \centering
         \includegraphics[width=\textwidth]{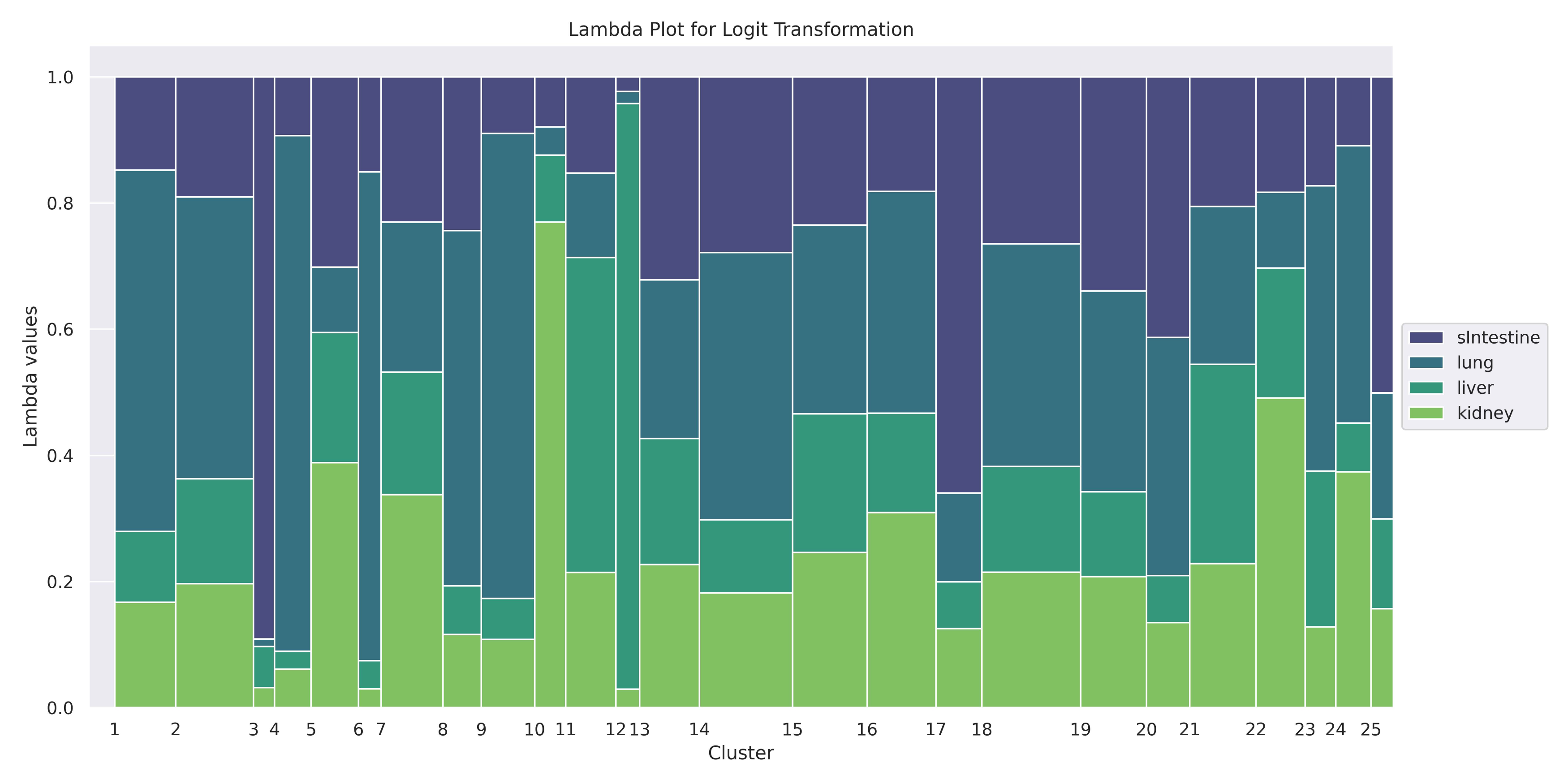}
         \caption{Cluster behavior for the prostate cancer cell line dataset:Transformation Method}  \label{fig:Transformation_mouse}
     \end{subfigure}
        \caption{Three cluster visualization plots for the mouse tissue dataset}
        \label{fig:Mouse}
\end{figure}

As is well known, visualizing results of a co-expression analysis for RNA-seq data can be rather complicated. This is due to the extremely large dynamic range of digital gene expression and the fact that the more highly expressed genes tend to exhibit much higher variability than weakly expressed genes. The most appropriate manner in which the results of a co-expression analysis for RNA-seq data should be displayed is still an open research question. In this manuscript, we follow the visualization approach that is conceptually similar to what has been suggested first in \cite{rau2015co}. In this approach, bar widths correspond to the estimated  proportions for the corresponding cluster $\hat\pi_{k}.$ The proportion of reads that is attributed to each cell line in each cluster is represented by the corresponding colored segment within each bar.  More specifically, let $y_{ijq}$ be the read of the $i$th gene for the $q$th replicate of the $j$th condition where $j=1,\ldots,J$, $i=1,\ldots,I_{k}$, $q=1,\ldots,3$ and $I_{k}$ is the number of genes in the $k$th cluster. Then, the height of the vertical bin that corresponds to the $j$th condition in $k$th cluster is
\[
\lambda_{jk}=\frac{\sum_{i=1}^{i_{k}}\sum_{q=1}^{3}y_{ijq}}{\sum_{k=1}^{K}\sum_{i=1}^{i_{k}}\sum_{q=1}^{3}y_{ijq}}.
\]
This approach to visualization allows us to assess the relative size of clusters with different level of expression in different tissues. For example,  in different clusters rather easily. For example, one can see that clusters that contain mostly genes that are strongly expressed in small intestine and lungs tend to be larger than those that contain mostly genes that are expressed in liver or kidneys. The visualization results are presented in the Figure \eqref{fig:Mouse}.

In order to determine the biological relevance of our clustering results, we performed gene ontology (GO) enrichment analyses on genes in clusters.  This analysis identifies enrichment of biological processes amongst genes in clusters, thus investigating whether genes that our methods identify as co-expressed encode proteins that perform similar biological functions.    Initially, methods were applied to the mouse dataset to identify co-expression clusters.  Out of the $17$ clusters identified by the npMSL method, $16$ of them ($94.1$\%) show significant enrichment of biological processes.  The clusters identified are associated with highly specific  biological processes.  For example, cluster $2$ involves genes that function in water homeostasis (GO:0030104).  Cluster $12$ is clearly associated with processes involved in the adaptive immune response (GO:0002250), and also includes specific terms such as regulation of T cell activation (GO:0042110) and B cell activation (GO:0042113).  Cluster $16$ on the other hand involves GO terms related to the innate immune response, such as the innate immune response in mucosa (GO:0002227) and the defense response to Gram-positive bacterium (GO:0050830).  For this dataset, all three methods produce similar results: the Poisson method produces $19$ clusters that are all associated with statistically enriched GO terms while the logit transformation method results in the lowest performance, with $22$ out of $25$ clusters ($88.0$\%) enriched for biological processes.

\subsection{Human prostate cancer cell line dataset}
Again, we begin with the description of the dataset. We have four cell lines used that consist of samples that are sensitive to chemotherapeutic agents (C4-2 cells and LNCaP cells) as well as those that are resistant (MR49F and C4-2B cells). More specifically, a line of C4-2 cells is a subline of LNCaP cells (described below). Second, C4-2B cells are enzalutamide-resistant cells derived from C4-2 cells.  Third, LNCaP (Lymph Node Carcinoma of the Prostate) cell line is the line that has been established from a metastatic lesion of human prostatic adenocarcinoma. Finally, the MR49F cell line consists of enzalutamide-resistant cells derived from LNCaP cells. The resistant cell lines are of interest because these cells do not respond well to treatment.  Also, each of the cell lines has three replicates.  Filtering at the level of $1.5$ CPM as described above results in a file containing sequencing data for $16,247$ genes. As before, we also applied the log transform to the data as a first step in our data analysis. 

Note that it is unlikely that replicates in a cell line dataset are independent. If so, this dataset may violate a conditional independence model assumption that underlies the npMSL method. Due to this, an attempt to run our method on such a dataset may be viewed as a test of the algorithm's robustness to the violation of the conditional independence assumption.  Indeed, we discovered that the algorithm tends not to converge if the dataset is used ``as is" even when multiple choices of  starting values are considered. Therefore, we decided to treat zeros as missing observations and use a simple imputation procedure, which turned out to be useful. More specifically, every zero was substituted with a number generated from a uniform distribution on an interval $[0,A]$ where $A$ was the smallest count observed in the entire dataset. We would like to note here that the practice of treating zeros as missing observations, although not very common in the analysis of RNA-seq data, is quite widespread when analyzing the scRNA-seq (Single Cell RNA-seq) data \cite{hou2020systematic}. 

As for the previous dataset, we fit several Gaussian repeated measures models over a range of cluster numbers from $1$ to $20$. Once again, the comparison of the four model selection criteria suggests, using the majority vote criterion, that $17$ is an optimal number of clusters for the npMSL method. In this case, the procedure was not sensitive to the choice of starting values.

To see if a solution with a larger number of clusters may also be possible, we also tried to run this procedure over the range of $1$ to $30$ clusters. In this case, selection of a starting point turns out to be a difficult problem since many choices result in singularities. To avoid this problem, we ran a small-EM type initialization procedure using the repnormmixmodel.sel procedure from the R mixtools package \cite{benaglia2009mixtools}. The concept of small-EM type initialization procedure is described in detail in Section 2.3.2 of \cite{rau2015co}; see also \cite{biernacki2003choosing} for additional information. In our case, the only two differences were the use of the repeated measures model and the use of only $5$ iterations of the EM algorithm. With this initialization procedure, we found a set of starting values that did not result in a singularity; however, the suggested solution had only $11$ clusters that were not well separated. In the final analysis, we decided to keep the $17$ cluster solution that we described earlier as an optimal one.  

\begin{figure}
\centering
     \begin{subfigure}[t]{0.3\textwidth}
         \centering
         \includegraphics[width=\textwidth]{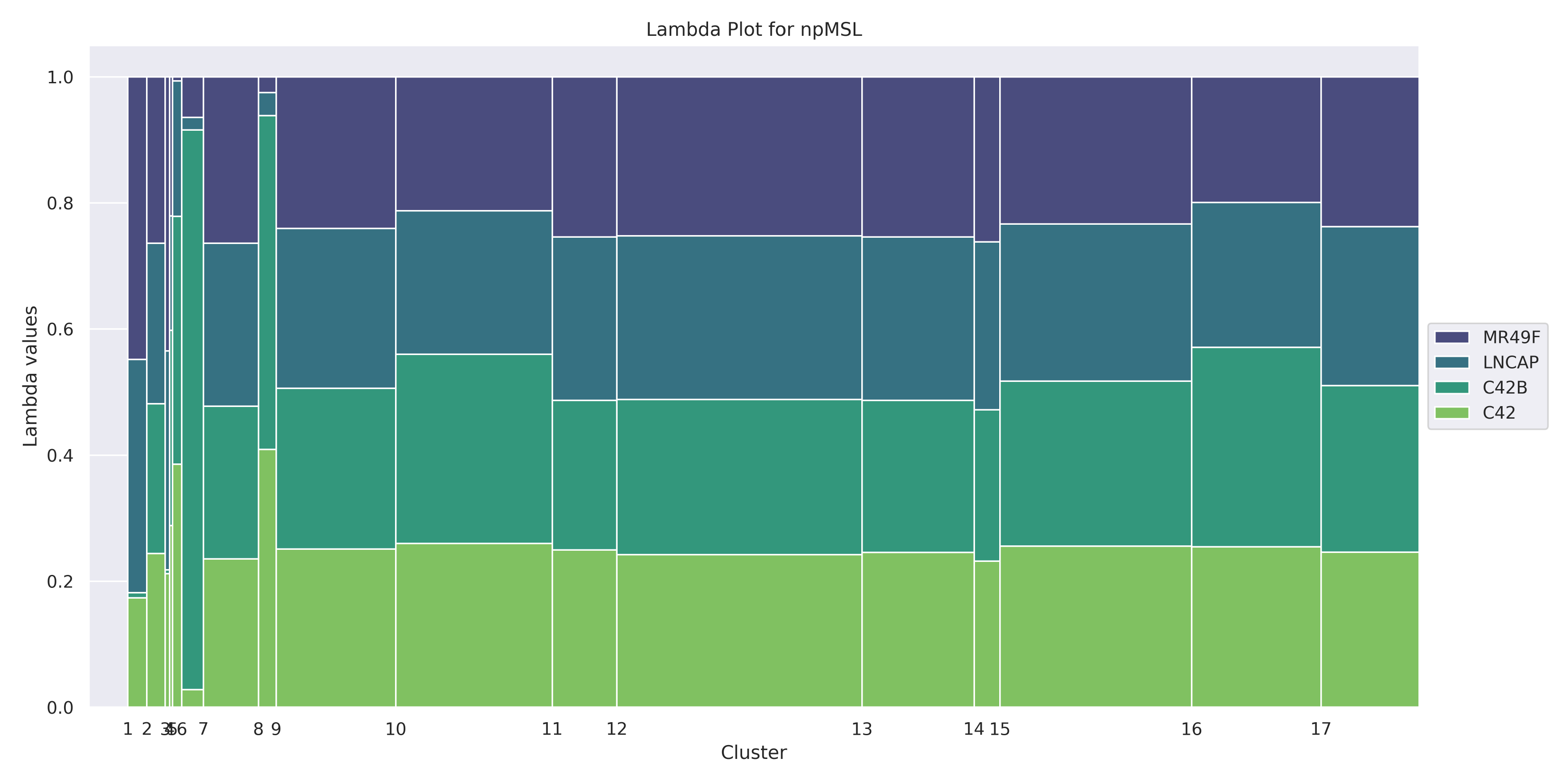}
         \caption{Cluster behavior for the prostate cancer cell line dataset:npMSL}
         \label{fig:npMSL}
     \end{subfigure}
    \hfill
\begin{subfigure}[t]{0.3\textwidth}
         \centering
         \includegraphics[width=\textwidth]{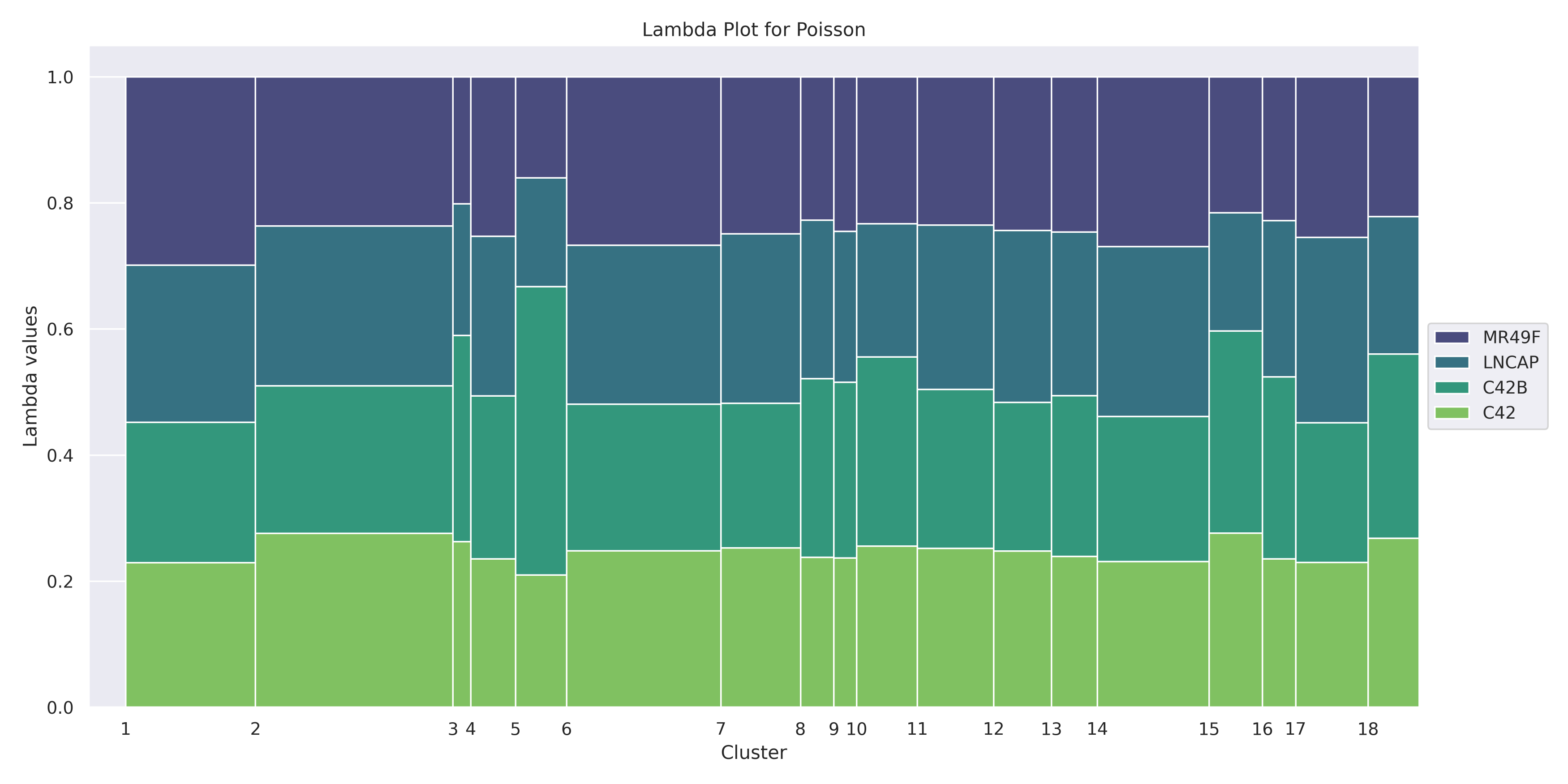}
         \caption{Cluster behavior for the prostate cancer cell line dataset:Poisson method}
         \label{fig:Poisson}
     \end{subfigure}
     \hfill
\begin{subfigure}[t]{0.3\textwidth}
         \centering
         \includegraphics[width=\textwidth]{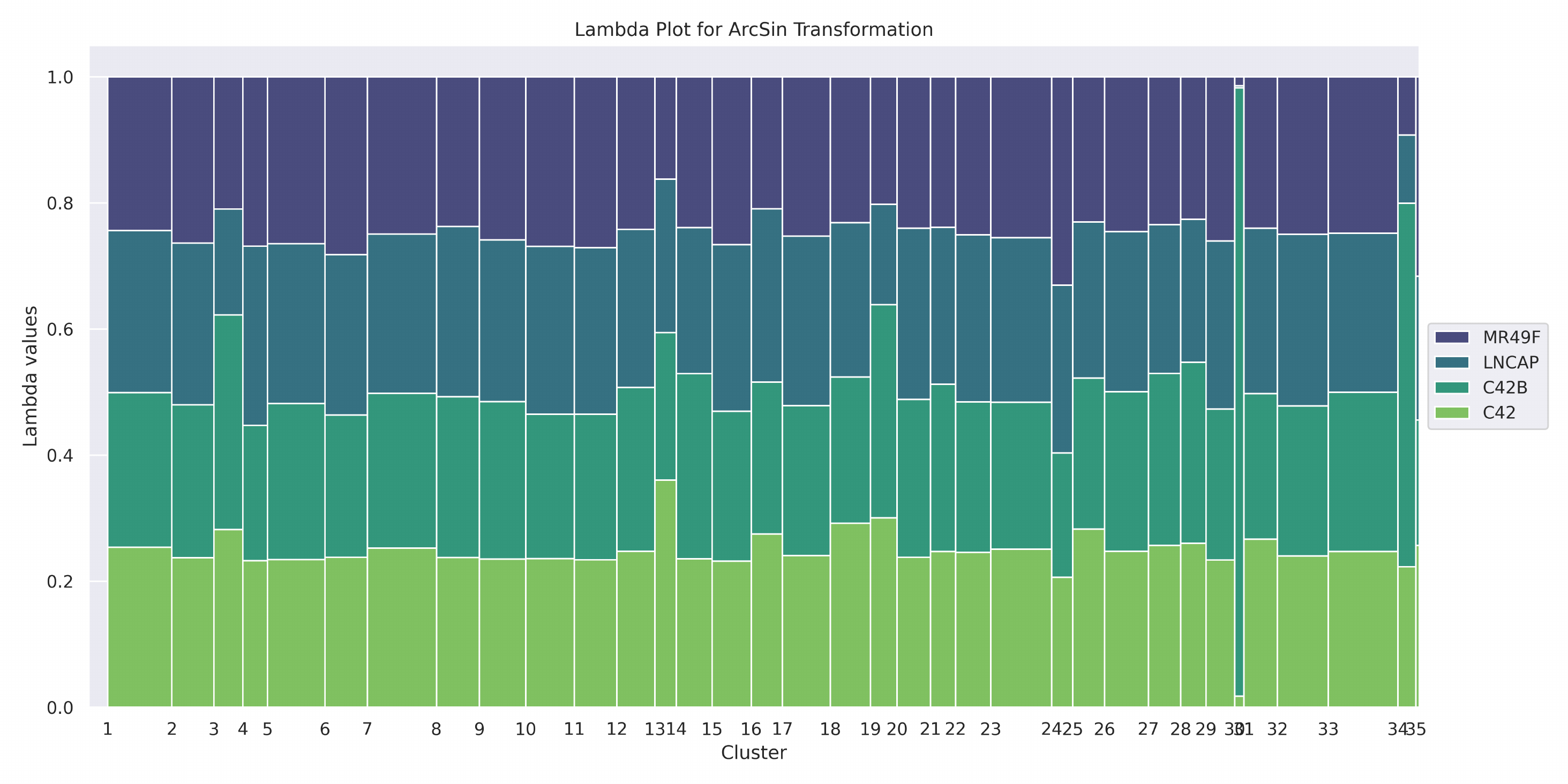}
         \caption{Cluster behavior for the prostate cancer cell line dataset:Transformation Method}  \label{fig:Transformation}
     \end{subfigure}
        \caption{Three cluster visualization plots for the prostate cancer cell line dataset}
        \label{fig:Cell_Line}
\end{figure}

For visualization purposes, we followed the same approach as for the previous dataset. The results are given in the Figure \eqref{fig:Cell_Line}.


For comparison purposes, we compare clustering results obtained using the npMSL method with two other methods commonly used for model-based clustering of RNA-seq data. The first of these methods is the Poisson method of \cite{rau2015co} while the second one is the transformation-based method of \cite{rau2018transformation}. We believe that the Poisson method is an appropriate comparison benchmark since it, alongside our methods, assumes conditional independence of marginals. Our method simply takes this assumption one step further and does not impose any specific distributional assumption on the marginal distributions. Note that the Poisson distribution only models the integer-valued data. Thus, when applying the Poisson method, we have used the raw data counts instead of the FPKM normalized dataset. We used the so-called slope heuristics method to select the optimal number of clusters in this case; our approach resulted in the choice of $18$ clusters when running the slope heuristics approach over a range from $1$ to $35$ clusters. The slope heuristics method has been suggested in \cite{birge2001gaussian,birge2007minimal} while some of the practical suggestions concerning its use can be found in \cite{baudry2012slope}. 

Another method that we used for comparison purposes has been the transformation method of \cite{rau2018transformation}. This method is based on the use of data transformations in conjunction with Gaussian mixture models. It also uses a penalized model selection criterion to select the number of clusters present in the data. Compared to the Poisson method, it allows for modeling of per-cluster correlation among biological samples. To analyze our dataset, we used this method with an arcsin transform. The number of clusters was selected using the so-called ICL (Integrated Complete Likelihood) method \cite{biernacki2000assessing} which is a type of the penalized selection criterion method. Using this criterion over a possible range of $1$ to $40$ clusters, we found that $2$5 seemed to be the optimal number in this case.

All three clustering methods identify fewer biologically relevant clusters when applied to this cell line dataset compared to the mouse dataset. Furthermore, many of the enriched biological processes identified are at a higher level and are thus less specific as compared to those identified in the mouse dataset.  Overall, the npMSL method identifies more biologically relevant clusters than either the Poisson or the arcSin transformation methods.  Out of $17$ clusters identified by the npMSL method, $14$ ($82.4$\% of clusters) show enrichment of biological process GO terms.  For example, cluster $1$ is associated with terms related to translation initiation (GO:0006413) such as cotranslational protein targeting to membrane (GO:0006613) and SRP-dependent cotranslational protein targeting to membrane (GO:0006614).  Cluster $6$ is associated with non-coding RNA processing (GO:0034470).  Cluster $10$ is heavily enriched for mRNA catabolic processes(GO:0006402) and cluster $14$ is involved in tRNA processing (GO:0008033) including tRNA modifications (GO:0006400) and tRNA metabolic processes (GO:0006399).  In comparison, the arcSin transformation method resulted in $12$ out of $35$ clusters ($34.2$\% of the clusters) showing a significant enrichment of GO terms and the clusters derived from the Poisson method results in only $5$ out of $18$ clusters ($27.8$\% of clusters) with enrichment of GO terms.   

\section{Discussion}

In this manuscript, we proposed a possible method of discovering gene co-expression networks in RNA seq data. The suggested method entails the use of a rigorous framework for parameter estimation, based on MM (Maximization-Minorization) algorithms. The model we use is distinct from most models routinely used in the clustering of digital gene expression profiles since it is a nonparametric one. More specifically, we assume that, conditional on knowing the cluster an observations has been generated from, biological samples (corresponding to marginal distributions)  are independent. Moreover, each marginal distribution is not assumed to belong to any predetermined distributional family.  


Conceptually, our proposal is closest to that of \cite{rau2015co} that also uses model with conditionally independent biological samples but imposes a Poisson restriction on marginal distributions. The Poisson assumption may not be very realistic in practice since it does not account for the overdispersion routinely observed in the RNA-seq data. The attempts to handle the overdispersion problem from the parametric viewpoint typically concentrate on proposing specific distribution to handle it such as e.g. negative binomial in \cite{si2014model} or the multivariate Poisson-lognormal in \cite{subedi2020parsimonious}. By comparison, our method is a more general one since it imposes no specific assumptions on the marginal distributions of individual biological samples. Moreover, fitting per-gene dispersion parameters using the experimental data with multiple conditions is typically rather difficult due to a small number of replicates available in such datasets. Our proposed method avoids this problem altogether by avoiding the parametric framework for marginal distributions. The use of nonparametric multivariate clustering in bioinformatics has so far been extremely limited \cite{anchang2014ccast}; to the best of our knowledge, these models have not been used for the discovery of gene co-expression networks before. 

The algorithm that we use to fit the suggested model has been described earlier in the statistical literature \cite{levine2011maximum} and its use illustrated in various simulation settings. Moreover, there does not seem to be a general agreement on how to simulate the data resembling the true RNA-seq data in an optimal way. Due to this, we do not include a simulation study in this manuscript. Instead, we simply demonstrate the behavior of our method on two real datasets: a mouse tissue dataset and a human prostate cancer cell line dataset.  The results are compared to those obtained by applying the Poisson method of \cite{rau2015co} and a transformation method of \cite{rau2018transformation} to those same datasets. In both cases, the npMSL method seems to identify a number of biologically meaningful clusters that is at least comparable to the number produced by other methods; in the human prostate cancer cell-line dataset case, it outperforms both alternative methods. For both datasets, the performance of the methods has also been compared using the adjusted Rand index (ARI) \cite{hubert1985comparing}. 
The results are summarized in Tables \eqref{Table1} and \eqref{Table2}. The most salient feature of this summaries is that the clustering produced by the nonparametric npMSL method is very different from those produced by either Poisson method or a transformation-based method. In other words, these seem to represent different solutions of a problem. We would like to note here that in the only case we are aware of where a researcher tried to applied a nonparametric method to clustering of the RNA-seq data the conclusion was a rather similar one - the clustering produced has been quite different from those produced by parametric methods \cite{mazo2019constraining}.

\begin{table}
\centering
\begin{tabular}{|c|c|c|c|} 
 \hline
 & npMSL & Poisson & Transformation \\ 
 \hline
 npMSL & 1 & 0.123015052
 & 0.089651457
 \\ 
 \hline
 Poisson &0.123015052
  &1 &  0.22480933
\\ 
 \hline
 Transformation &0.089651457
 & 0.22480933
 &1 \\
 \hline
\end{tabular}
\caption{ARI values for the mouse tissue dataset}
\label{Table1}
\end{table}

\begin{table}
\centering
\begin{tabular}{|c|c|c|c|} 
 \hline
 & npMSL & Poisson & Transformation \\ 
 \hline
 npMSL & 1 &0.013246073
 & 0.026826513
 \\ 
 \hline
 Poisson &0.013246073
  &1 &  0.110776115
\\ 
 \hline
 Transformation &0.026826513
 &0.110776115
 &1 \\
 \hline
\end{tabular}
\caption{ARI values for the human prostate cancer cell line dataset}
\label{Table2}
\end{table}


The suggested model does not take into account possible dependence between biological samples because it enforces the conditional independence between them. The natural next step in this line of research is to try to lift this assumption while preserving the general nonparametric nature of marginal distributions in our proposed model. One way this can be done is by considering explicit dependence structures between biological samples modeled using the so-called copula functions \cite{mazo2019constraining}. A useful direction of the future research will be the introduction of specific multivariate copulas such as e.g. Gaussian copula to model the dependence between biological samples in RNA-seq data. Note that this approach allows a researcher to model any type of dependence and not just correlation which is often claimed to be the benefit of using certain so-called hidden layer approaches \cite{silva2019multivariate}. Our research in this direction is ongoing. 





\section{Funding}
This work was supported in part by funds from the Purdue University Center for Cancer Research[P30CA082709]; the IU Comprehensive Cancer Center [P30ca082709]; and the Walther Cancer Foundation.

\bibliographystyle{unsrt}
\bibliography{bibliography_NAL}

\end{document}